\documentstyle[epsfig,longtable]{aipproc}

\begin{document}
\title{Hydrodynamics and Radiation\\ from a Relativistic Expanding Jet\\
 with Applications to GRB Afterglow}

\author{Jonathan Granot$^*$, Mark Miller$^{\dagger}$, Tsvi Piran$^*$,
and Wai-Mo Suen$^{\dagger}$}

\address{$^*$Racah Institute, Hebrew University, Jerusalem 91904, Israel\\
$^{\dagger}$Department of Physics, Washington University, St. Luis,
MO, USA}

\maketitle

\begin{abstract}
  We describe fully relativistic three dimensional calculations of the
  slowing down and spreading of a relativistic jet by an external
  medium like the ISM. We calculate the synchrotron spectra and light
  curves using the conditions determined by the hydrodynamic
  calculations.  Preliminary results with a moderate resolution are
  presented here.  Higher resolution calculations are in progress.
\end{abstract}

\section*{Introduction}
The level of beaming in GRBs is one of the most interesting open
questions in this subject.  The relativistic flow which drives a GRB
may range from isotropic to strongly collimated into a narrow opening
angle. The degree of collimation (beaming) of the outflow has many
implications on most aspects GRB research, such as the requirements
from the ``inner engine'', the energy release and the event rate.
During the prompt emission the Lorentz factor of the flow is very high
($\gamma>100$), and due to relativistic aberration of light, only a
narrow angle of $\sim 1/\gamma$ around the line of sight (LOS) is
visible. During the afterglow stage, the flow decelerates and an
increasingly larger angle becomes visible. As long as
$\gamma>1/\theta$ an outflow collimated within an angle $\theta$
around the LOS produces the same observed radiation as if it were part
of a spherical outflow. Once $\gamma$ drops below $1/\theta$, the
observer can notice that radiation arrives only from within an angle
$\theta$ around the LOS, instead of $1/\gamma$ as in the spherical
case.  Sideways expansion is also expected to become important when
$\gamma\sim 1/\theta$ \cite{Rhoads,SPH}. These two effects combine to
create a break in the light curve at $\gamma\sim 1/\theta$, but it is
not quite clear whether this break is sharp enough to be detected
\cite{Rhoads,SPH,WL}. To explore this question we have performed fully
relativistic three dimensional simulations that follow the slowing
down and the lateral expansion of a relativistic jet. We then
calculate the synchrotron light curve and spectrum using the conditions
determined by the hydrodynamical simulation.
 
\section*{The Hydrodynamics}

We use a fully relativistic three dimensional code for the
hydrodynamical calculations. The initial conditions 
are a wedge with an  opening angle $\theta_0=0.2$ taken from the
Blandford McKee \cite{BM}, BM hereafter, self similar spherical
solution, embedded in a cold and homogeneous ambient medium. For this
initial opening angle, the jet is expected to show considerable
lateral expansion when $\gamma\sim 1/\theta_0=5$, where $\gamma$ is
the Lorentz factor of the fluid. The BM solution used for the initial
conditions was therefore at the time when $\Gamma=10$, where
$\Gamma=\sqrt{2}\gamma$ is the Lorentz factor of the shock. The 
total isotropic energy was $E=10^{52}\ {\rm ergs}$, and the ambient number
density was $n=1\ {\rm cm^{-3}}$.

\begin{figure}
\centering
\noindent
\includegraphics[width=14cm]{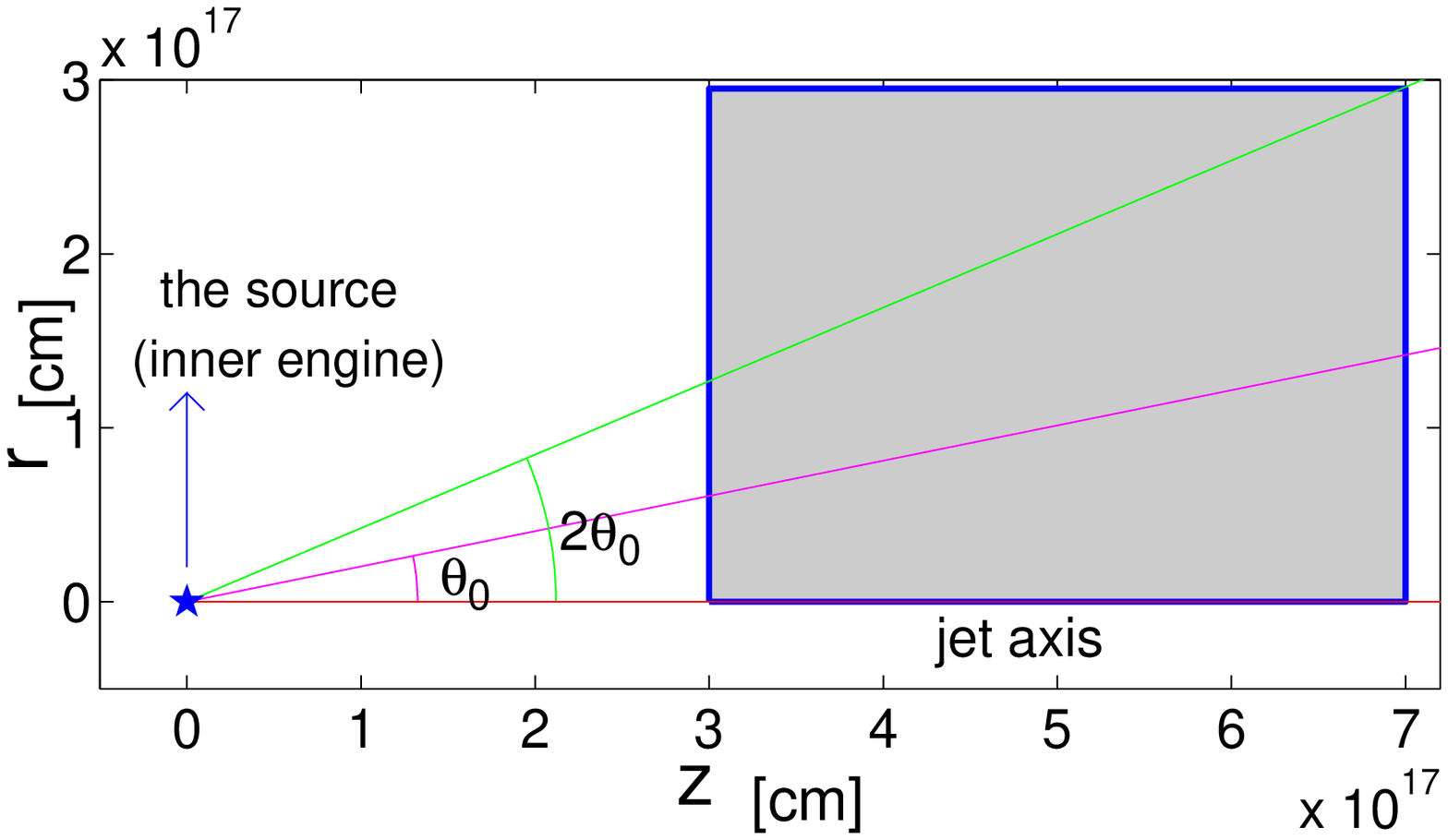}
\vspace{10pt}
\caption{\label{Fig1}The region depicted
  in Figures (\ref{Fig2}-\ref{Fig5}) is within the shaded region,
  outlined in blue.  $z$ and $r$ are the cylindrical coordinates.
  The jet axis is in the $z$ direction. Two colored lines which
  indicate opening angles of $\theta_0$ and $2\theta_0$ are added to
  help follow the lateral expansion.}
\end{figure}

In Figures \ref{Fig2}-\ref{Fig5} we present snapshots of  
the number density, internal energy density, Lorentz
factor and velocity field of the fluid, as the jet slows down and
spreads sideways. An explanation of what is seen in these snapshots is
given in Figure \ref{Fig1}. The snapshots are taken at consecutive
times, ranging $t=137-282\ {\rm days}$ in the rest frame of the
ambient medium (corresponding roughly to observer times from one and a
half to twenty days for an observer along the jet axis). The results
shown in this work are still preliminary. The resolution is 
not sufficient to resolve  the very thin initial shell.
For this reason the maximal Lorentz factor of the matter is just over
$3$, instead of $10/\sqrt{2}\cong 7.07$ (see Figure \ref{Fig4}).
Higher resolution runs are in progress.

\section*{Calculating Light Curves and Spectra}

The local emission coefficient at a given space-time point is
calculated directly from the hydrodynamical quantities there.
The magnetic field and electron
energy densities are assumed to hold constant fractions, $\epsilon_B$
and $\epsilon_e$, respectively, of the internal energy, while the
electrons posses a power law energy distribution,
$N(\gamma_e)\propto\gamma_e^{-p}$. The local emissivity is
approximated by a broken power law: $F_{\nu}\propto \nu^{1/3}$ and $\ 
\nu^{(1-p)/2}$, below and above the typical synchrotron frequency,
respectively \cite{GPS}.

The light curves and spectra are calculated for several viewing angles
with respect to the jet axis. Once the emission coefficient is
determined in the local frame, it is transformed to the frame of each
observer. The time of arrival to each observer is then calculated,
and the contributions are sumed over space-time,  producing the
various light curves. This is done for several frequencies,
simultaneously, so that the spectrum may be obtained, as well as the
light curve at different frequencies.

A few light curves, calculated for $\epsilon_B=\epsilon_e=0.1$ and
$p=2.5$, are shown in Figure (\ref{Fig6}). These light curves 
serve to demonstrate the potential of this approach.
Future simulations are expected to achieve sufficient resolution to
produce realistic light curves which could be compared with afterglow
observations.

\newpage

\begin{figure}
\caption{\label{Fig2} The proper number density, in ${\rm cm^{-3}}$.}
\end{figure}
\begin{figure}
\caption{\label{Fig3}
The internal energy density, in units of $m_pc^2/{\rm cm^3}$.}
\end{figure}
\begin{figure}
\caption{\label{Fig4}The bulk Lorentz factor of the fluid.}
\end{figure}

\clearpage

\begin{figure}
\centering
\noindent
\includegraphics[width=4cm]{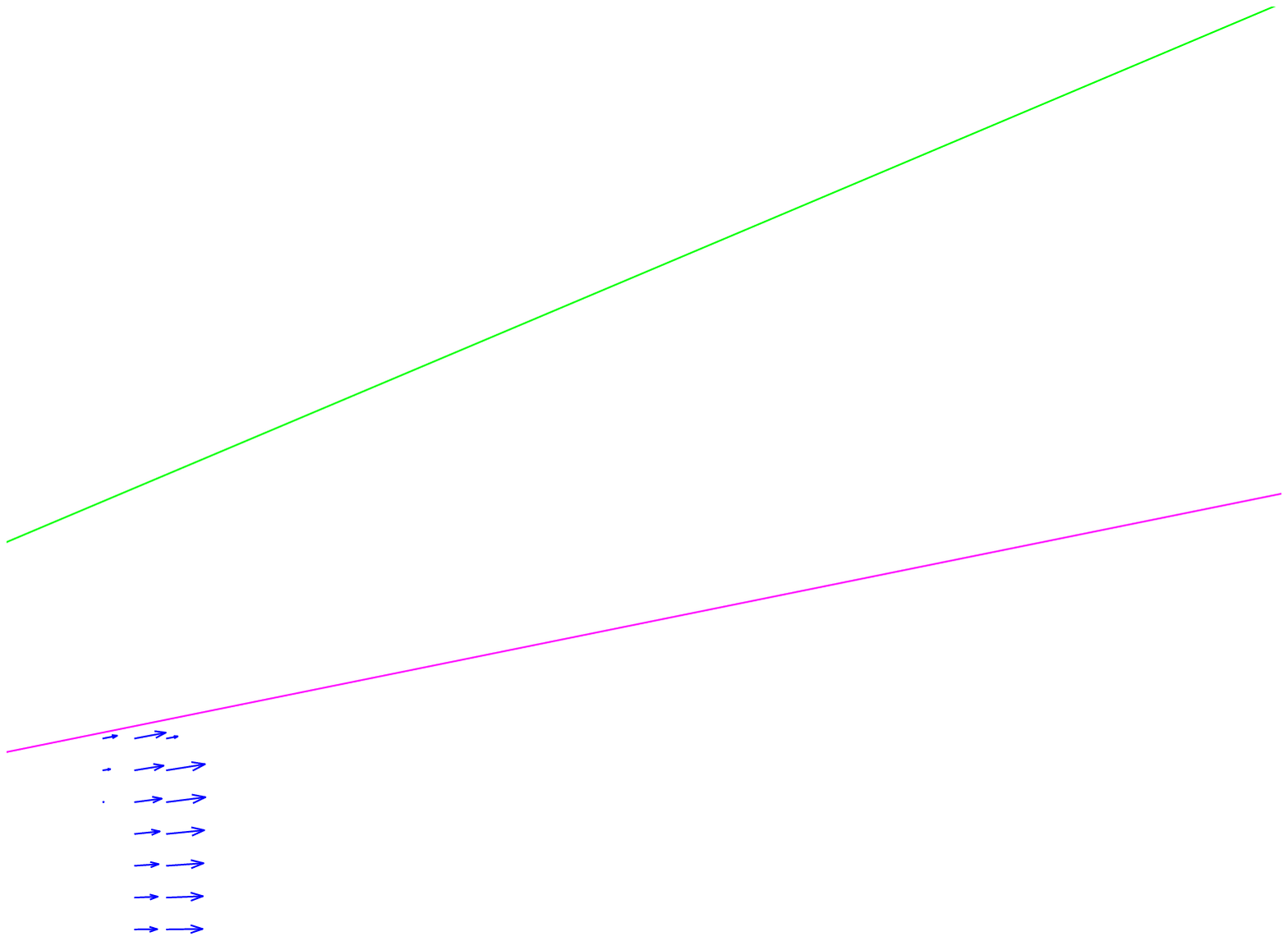}
\hspace{5pt}
\includegraphics[width=4cm]{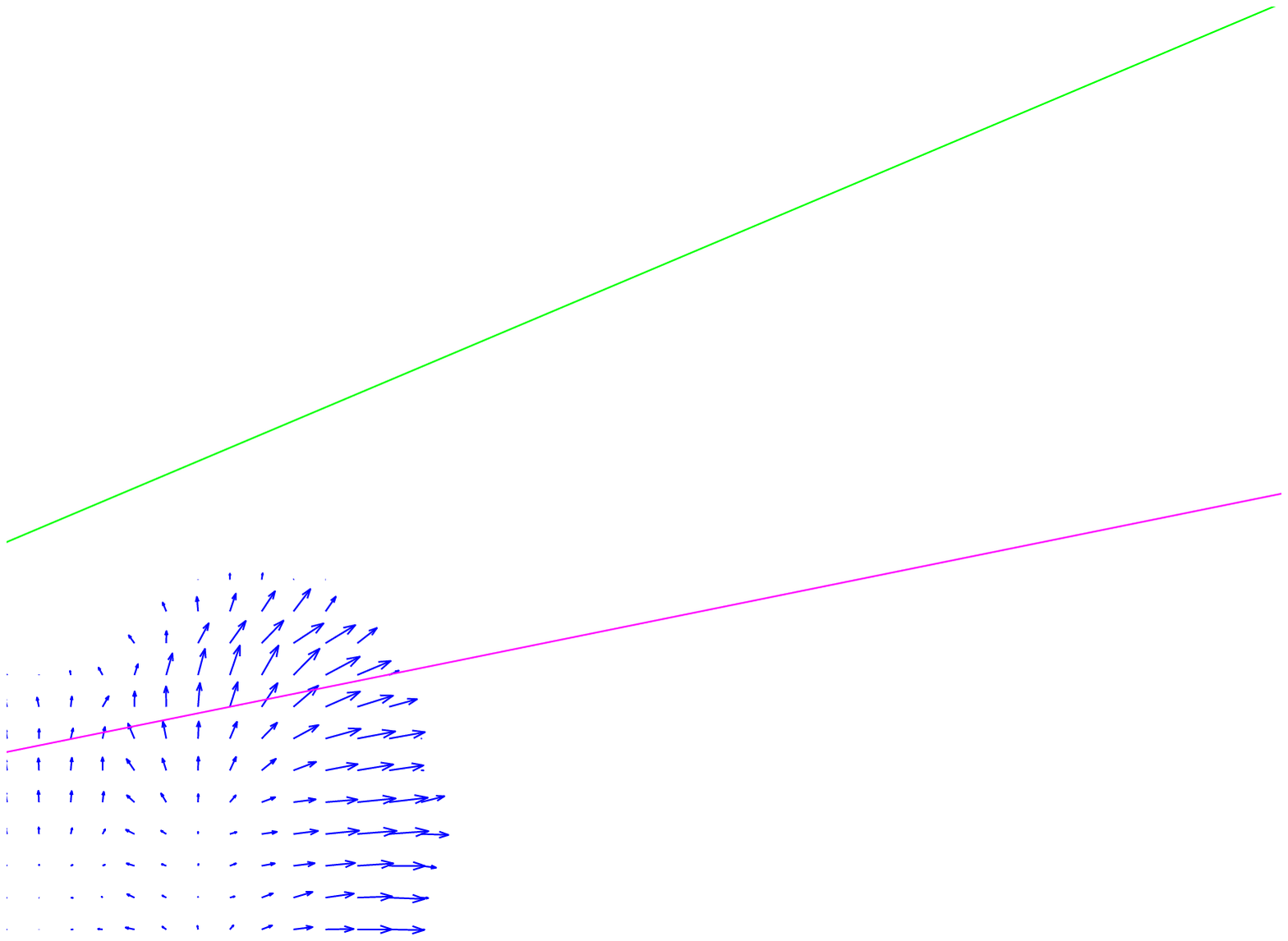}
\hspace{5pt}
\includegraphics[width=4cm]{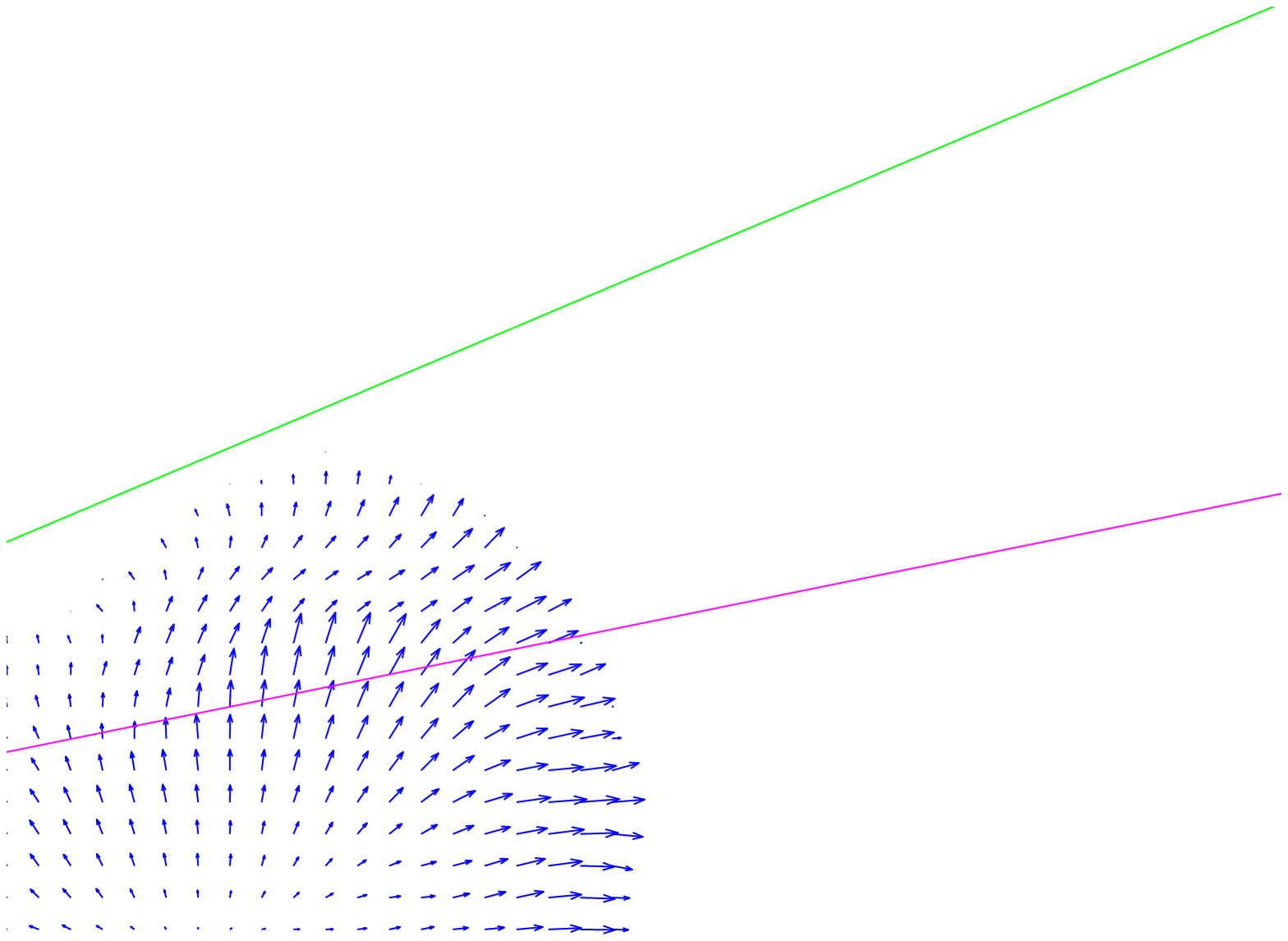}

\vspace{10pt}

\includegraphics[width=4cm]{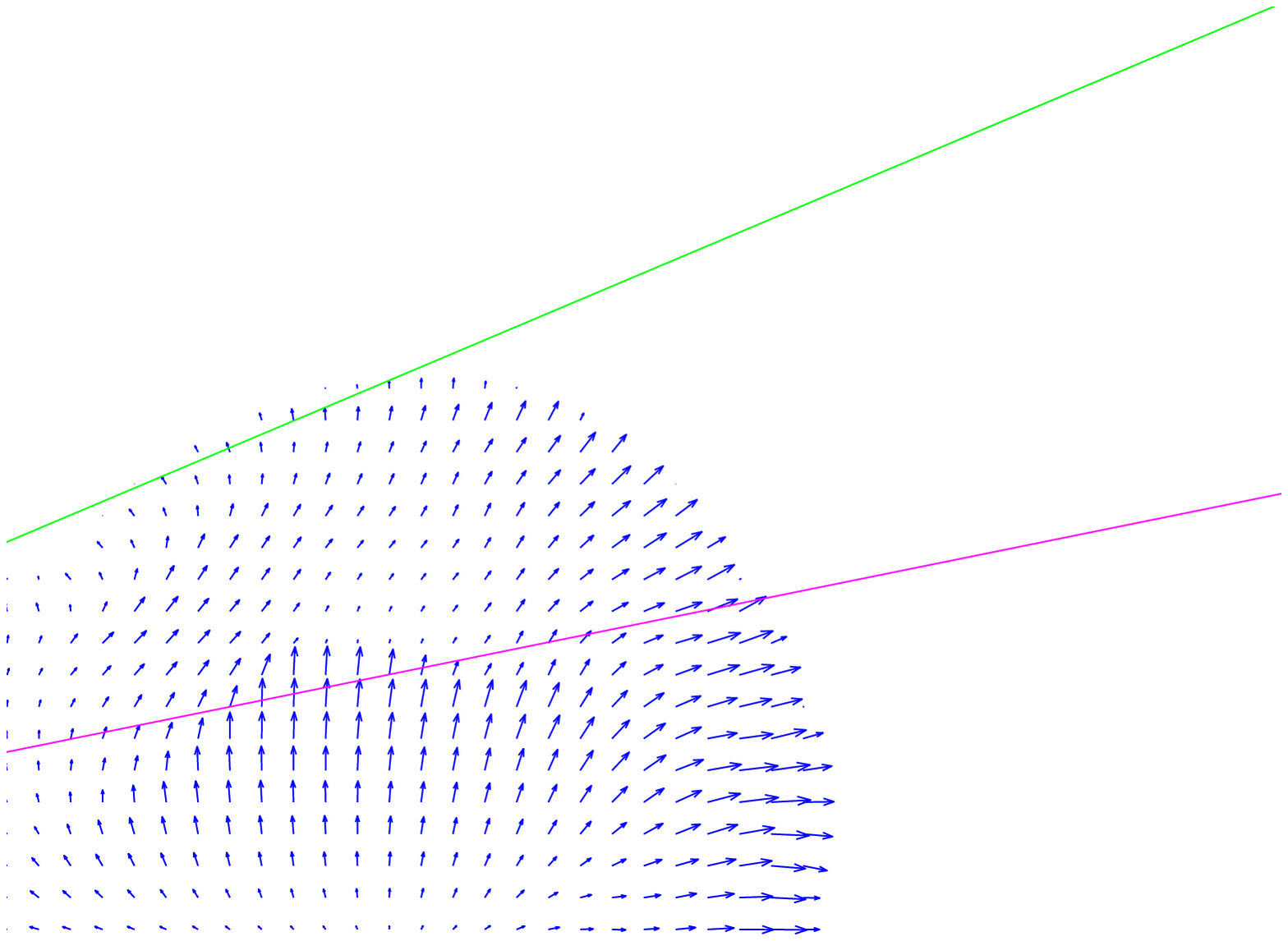}
\hspace{5pt}
\includegraphics[width=4cm]{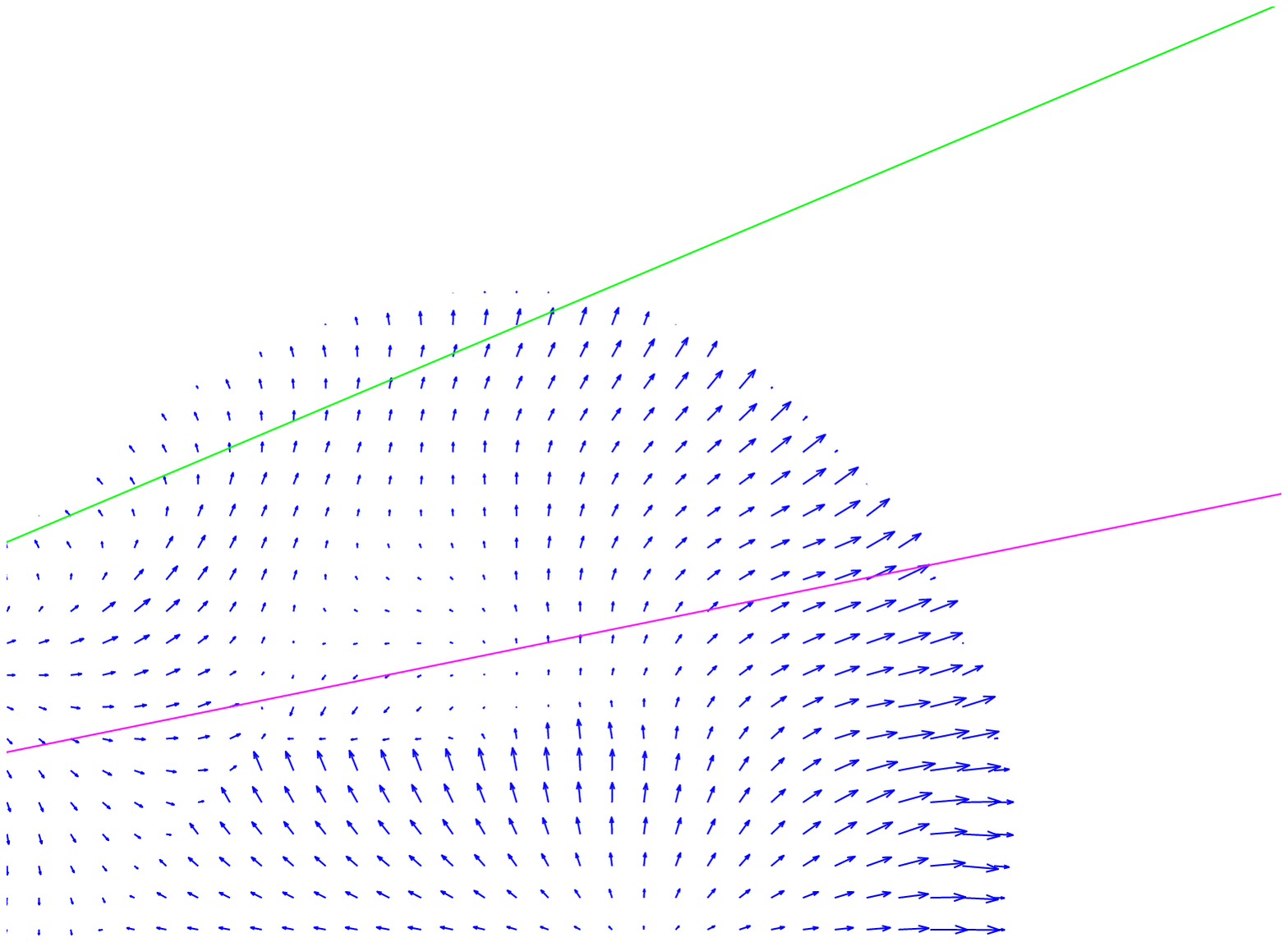}
\hspace{5pt}
\includegraphics[width=4cm]{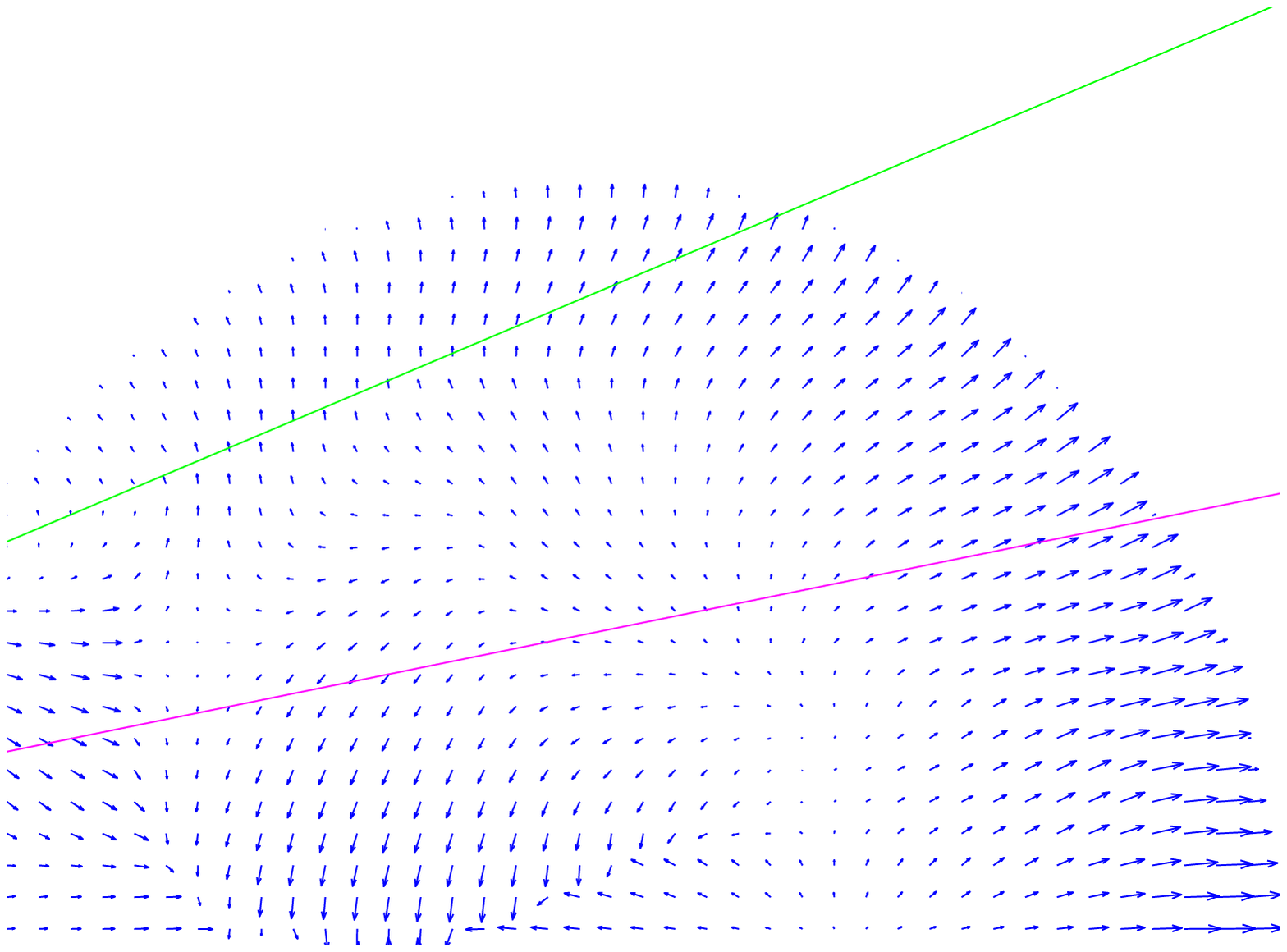}
\\
\vspace{10pt}
\caption{\label{Fig5}The velocity field of the fluid.
  Each arrow points at the direction of the velocity at the point
  where it begins, while its length is proportional to the size of the
  velocity.}
\end{figure}

\clearpage

\begin{figure}
\centering
\noindent
\includegraphics[width=8cm]{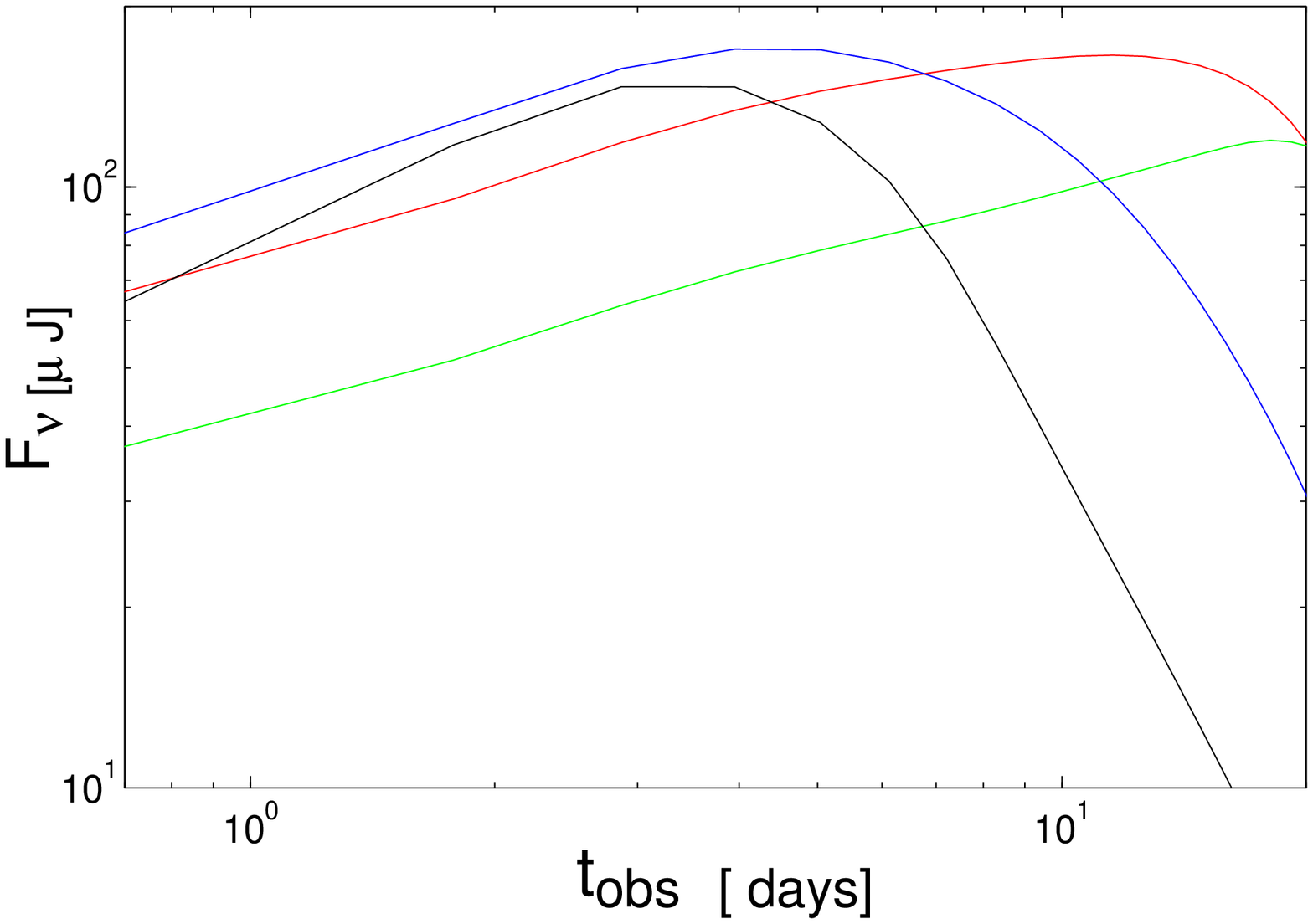}
\vspace{10pt}
\caption{\label{Fig6}The green, red, blue and
  black curves represent the light curves at the observed frequencies
  $10^8, 10^9, 10^{10}$ and $10^{11}\ {\rm Hz}$, respectively, as seen
  by an observer at a distance of $10^{28}\ {\rm cm}$ along the jet
  axis.}
\end{figure}


\begin{references}
\bibitem{BM}Blandford, R.D. \& McKee, C.F. 1976, Phys. of fluids, 
{\bf 19}, 1130.  
\bibitem{GPS}Granot, J, Piran, T. \& Sari, R. 1999,
  ApJ, {\bf 513}, 679.
\bibitem{Rhoads}Rhoads, J.E., 1999, ApJ submitted, astro-ph9903399.
\bibitem{SPH}Sari, R., Piran, T. \& Halpren, J.P., 1999, ApJ, {\bf 519}, L17.
\bibitem{WL}Wei, D.M. \& Lu, T. 1999, astro-ph9908273
\end{references}
\end{document}